\documentclass[aps,preprintnumbers,twocolumn,nofootinbib,amsmath,amssymb,showpacs]{revtex4}

\usepackage{amsmath,amssymb}
\usepackage{graphicx}
\usepackage{color}

\def\vx{{\mathbf x}}

\newcommand{\pri}{\!\,^{\prime}}
\newcommand{\prii}{\!\!\!\!\!\,^{\prime}\,}

\begin{document}

\title{Casimir forces in a piston geometry
at zero and finite temperatures}

\author{M.~P.~Hertzberg,$^{1,2}$ R.~L.~Jaffe,$^{1,2}$ M.~Kardar,$^2$ and A.~Scardicchio$^{3}$}
\affiliation{$^1$Center for Theoretical Physics, Laboratory for Nuclear Science, and \\
$^2$Department of Physics, 
Massachusetts Institute of Technology, Cambridge, MA 02139, USA \\
$^3$ Princeton Center for Theoretical Physics and Department of Physics \\
Princeton University, Princeton, NJ 08544, USA}


\begin{abstract}
We study Casimir forces on the partition in a closed box (piston) with perfect metallic boundary conditions.
Related closed geometries have generated interest as candidates for a repulsive force.
By using an optical path expansion we solve exactly the case of a piston with a rectangular cross section, and find that the force always attracts the partition to the nearest base.
For arbitrary cross sections, we can use an expansion for the density of states to
compute the force in the limit of small height to width ratios.
The corrections to the force between parallel plates are found to have interesting dependence
on the shape of the cross section.
Finally, for temperatures in the range of experimental interest we compute finite temperature
corrections to the force (again assuming perfect boundaries).
\end{abstract}
\vspace*{-\bigskipamount} \preprint{MIT-CTP-3703}

\maketitle

\section{Introduction}
\label{intro}

A striking macroscopic manifestation of quantum electrodynamics is the attraction of {\it neutral} metals.
In 1948 Casimir predicted that such a force results from the modification of the ground state energy 
of the photon field due to the presence of conducting boundary conditions \cite{Casimir}.
The energy spectrum is modified in a fashion that depends on the separation between the plates, 
$a$. While the zero-point energy is itself infinite, its variation with $a$ gives rise to a finite force. 
High precision measurements, following the pioneering work of Lamoreaux in 1997 \cite{expLam}, 
have renewed interest in this subject.
A review of experimental attempts to measure the force prior to 1997, and the many improvements
since then, can be found in Ref.~\cite{Onofrio}.
As one example, we note experiments by Mohideen {\it et al.}\cite{expMoh}, using an atomic force microscope, which
have confirmed Casimir's prediction from 100nm to several $\mu$m, to a few percent accuracy.
Forces at these scales are relevant to operation of  micro-electromechanical systems (MEMS),
such as the actuator constructed by Chan {\it et al.} \cite{Chan} to control the frequency of 
oscillation of a nanodevice.
They also appear as an undesirable background in precision experiments such as those that test 
gravity at the sub-millimeter scale \cite{Adelberger}.  

An undesirable aspect of the Casimir attraction is that it can cause the collapse of a device,
a phenomenon known as ``stiction'' \cite{Serry}.
This has motivated the search for circumstances where the attractive force can be reduced,
or even made repulsive~\cite{KKMR}.
The Casimir force, of course, depends sensitively on shape, as evidenced from comparison of
known geometries from parallel plates, to the sphere opposite a plane \cite{Gies}, 
the cylinder opposite a plane \cite{Emig2}, eccentric cylinders \cite{Dalvit}, the
hyperboloid opposite a plane \cite{Oliver}, a grating \cite{Emig}, a corrugated plane \cite{Kardar}.
The possibility of a repulsive Casimir force  between perfect metals  can be traced
to a computation of energy of a spherical shell by Boyer \cite{Boyer}, who found that
the finite part of this energy is opposite in sign to that for parallel plates.
This term can be regarded as a positive pressure favoring an increased 
radius for the sphere,
{\em if it were the only consequence of changing the radius}.
The same sign is obtained for a square in 2-dimensions and a cube in 3 dimensions
\cite{group,Wolfram}.
For a parallelepiped with a square base of width $b$ and height $a$, the finite part of the Casimir
energy is positive for aspect ratios of $0.408<a/b<3.48$. 
This would again imply a repulsive force in this regime if there were no other energy contributions accompanying deformations at a fixed aspect ratio.
Of course, it is impossible to change the size of a material sphere (or cube) without changing its
surface area, and other contributions to its cohesive energy.  For example, a spherical shell cut into two equal hemispheres which are then separated has superficial resemblance to the Boyer calculation.  However the cut changes the geometry, and it can in fact be shown\cite{Klich} that the two hemispheres attract.

The piston geometry, first considered by Cavalcanti \cite{Caval} (in 2 dimensions)
and further considered in Refs.~\cite{Hertzberg,Val_Edery} (in 3 dimensions),  is closely related to the 
parallelepiped discussed above.\footnote{The piston geometry was earlier mentioned in Ref.~\cite{Power}.}
As depicted in  Fig.~\ref{picpist}, we examine a piston of height $h$, with a movable partition
at a distance $a$ from the lower base. 
The simplest case is that of a rectangular base, but this can be 
generalized to arbitrary cross sections. 
This set-up is experimentally realizable, and does not require any deformations
of the materials as the partition is moved.
The force resulting from rigid displacements of this piece is perfectly well defined,
and free from various ambiguities due to cut-offs and divergences that will be discussed later. 
In particular, we indeed find the finite part of the energy can be ``repulsive" if only one of the boxes
adjoining the partition is considered, while if both compartments are included, the net force on
the partition is attractive (in the sense that it is pulled to the closest base). 

This paper expands on a previous brief publication of our results \cite{Hertzberg},
and is organized as follows.
Section~\ref{prelims} introduces the technical tools preliminary to the calculations,
and includes sections on cutoffs and divergences, the optical path approach, and on
the decomposition of the electromagnetic (EM) field into two scalar field (transverse magnetic
and transverse electric) with Dirichlet and Neumann boundary conditions (respectively).
Details of the computation for pistons of rectangular cross section are presented in
Section~\ref{rectangle}, and the origin of the cancellations leading to a net attractive force
on the partition is discussed in some detail.
Interestingly, it is possible to provide results that are asymptotically exact in the limit
of small separations for cross sections of arbitrary shape. 
As discussed in Section~\ref{general}, there is an interesting dependence on the shape 
in this limit, related to the resolution with which the cross section is viewed.
Finally, in Section~\ref{thermal} we present new results pertaining to corrections to the 
Casimir force at finite temperatures in such closed geometries (for perfect metals).
We conclude with a brief summary (Section~\ref{conclusions}), and an Appendix.

\section{Preliminaries}
\label{prelims}

Before embarking on the calculation of the force on the partition, we introduce some relevant concepts
in this Section.
Subsection~\ref{cutoff} discusses the general structure of the divergences appearing in the calculation
of zero-point energies, and indirectly justifies our focus on the piston geometry.
The optical path approach, which is our computational method of choice is reviewed in
Sec.~\ref{optical}.
Another important aspect of the piston geometry is that it enables the decomposition of
the EM field into  Dirichlet and Neumann scalar fields, as presented in Sec.~\ref{eandm}.

\subsection{Cutoff dependence}
\label{cutoff}

Let us consider an empty cavity made of perfectly conducting material.
The Casimir energy of the EM field is a sum over the zero point energies of all modes compared to the energies in the absence of the material 
$E_{C}(\Lambda)=E(\Lambda)-E_{0}(\Lambda)= \frac{1}{2}\sum ^{\Lambda}\hbar \omega_m-\frac{1}{2}\sum ^{\Lambda}\hbar \omega^{0}_{m}$, 
and is divergent if the upper limit $\Lambda$ is taken to infinity. 
In a physical realization, the upper cutoff is roughly the plasma frequency of the metal, 
as it separates the modes that are reflected and those that are transmitted and are 
hence unaffected by the presence of the metallic boundaries. 
Based on general results for the density of states in a cavity \cite{Balian:1970fw}, 
we know $E(\Lambda)$ has an asymptotic form, with a leading term proportional
to the volume $V$ of the cavity and the fourth power of $\Lambda$ and sub-leading terms proportional to its surface area $S$, a length $L$ which is related to the average curvature of the walls (in a cavity with edges  but otherwise flat, like a parallelepiped, it is the total  length of the edges) proportional to 
$\Lambda^{3}$ and $\Lambda^{2}$ respectively, and so forth.
For example in the case of a scalar field with Dirichlet boundary conditions, we find
\begin{equation}
E(\Lambda)= \frac{3}{2\pi^2}V\Lambda^4 -\frac{1}{8\pi}S\Lambda^3+\frac{1}{32\pi}L\Lambda^2
+ ... +   \widetilde{E},
\label{eq0}
\end{equation}
where ``...'' denote lower order cutoff dependences,\footnote{{ 
For general geometries, there
are also linear and logarithmic terms in $\Lambda$, but for the class of geometries examined in this paper (pistons) there are no further terms in $\Lambda$.}} and $\widetilde{E}$ is the important finite part in the limit of $\Lambda\to\infty$.
 The EM field also enjoys a similar expansion, although some terms may be absent. 

Although the volume term is cancelled by an identical term in $E_{0}$, this is  not obviously the case for the other divergent terms (surface area, perimeter, and so on).
The energy of an isolated cavity is therefore
dependent on the physical properties of the metal.
A determination of the stresses in a single closed cavity 
requires detailed considerations of the metal, and its extrapolation to the
perfectly conducting limit will be problematic \cite{realistic}.
It is tempting to ignore these cutoff dependent terms, and to remove them in analogy to
the renormalization of ultraviolet divergences in quantum field theories. 
This is unjustified as there are no boundary counter-terms to cancel them, see Ref.~\cite{Jaffe}.
If however, we are interested in the force between rigid bodies, then any surface, perimeter, {\it etc.} terms
are independent of the distance between them, and a well defined (finite) force exists in the perfect conductor limit.

While the piston geometry considered in this paper is closely related to the parallelepiped cavities
considered in the literature, it does not suffer from problems associated with changes in shape. 
The overall volume, surface, and perimeter contributions are all unchanged as the height
of the partition is varied, and the force acting on it is finite and well defined.
The same observations led Cavalcanti~\cite{Caval} to consider a rectangular (2-dimensional) piston. 
He found that the force on the partition, though weakened relative to parallel lines, is attractive.

\subsection{Optical approach}
\label{optical}

The Casimir energy can be expressed as a sum over contributions of \emph{optical paths} \cite{Scard},
and much intuition into the problem is gained by classifying the corresponding paths.
For generic geometries this approach yields only an approximation to the exact result that ignores diffraction.  
Fortunately, it is exact for rectilinear geometries if reflections from edges and corners are properly included.  

Consider a free scalar field in spatial domain $\mathcal{D}$ obeying some prescribed boundary conditions (Dirichlet or Neumann) on the boundary $\mathcal{B}=\partial\mathcal{D}$.   The Casimir energy is defined as the sum over the zero point energies, $E = \sum \frac{1}{2}\hbar\omega$, where $\omega$ are the
eigenfrequencies in ${\mathcal D}$ (we refrain from subtracting $E_{0}$ for the moment).  This expression needs to be regularized, as explained in the previous section, by some cutoff $\Lambda$. We choose to implement this by a smoothing function $S_{\Lambda}(k)=e^{-k/\Lambda}$, and thus examine $E(\Lambda)=\sum_k\frac{1}{2}\hbar\omega(k) e^{-k/\Lambda}$.

The Casimir energy can be expressed in terms of the spectral Green's function $G({\bf x},{\bf x}',k)$ which satisfies the Helmholtz equation in $\mathcal{D}$ with a point source,
\begin{equation}
(\nabla'^2+k^2)G({\bf x}',{\bf x},k)=-\delta^3\!\left({\bf x}'-{\bf x}\right),
\end{equation}
and subject to the same boundary conditions on $\mathcal{B}$ as the field. 
The Casimir energy of a scalar field is then given by the integral over space and wavenumber of the imaginary part of $G$,
in the coincidence limit ${\bf x}'\to{\bf x}$ \cite{Jaffe} ($\hbar=c=1$), as
\begin{equation}
E(\Lambda)=\frac{1}{\pi}\Im\int_{0}^{\infty}dk\,k^2 e^{-k/\Lambda}\int_{\mathcal{D}}d{\bf x}\, G({\bf x},{\bf x},k)\,.
\label{density}
\end{equation}
The knowledge of the Helmholtz Greens function at coincident points allows us to calculate the Casimir energy of the configuration.

It is convenient to introduce a fictitious time $t$  and a corresponding
space-time propagator, $G({\bf x}',{\bf x},t)$,
defined as the Fourier transform of $G({\bf x}',{\bf x},k)$.
The propagator $G(t)$ can be expressed as
the functional integral of a free quantum particle of mass $m=1/2$ with appropriate phases associated with paths that reflect off the boundaries.

In the optical approach, the path integral is approximated as a sum over \emph{classical} paths of 
$\exp[i S_{p_r}({\bf x}',{\bf x},t)]$, weighted by
the Van Vleck determinant $D_r({\bf x}',{\bf x},t)$ \cite{Kurt}.
Here $S_{p_r}({\bf x}',{\bf x},t)$ is the classical action of a  path  $p_r$ 
from ${\bf x}$ to ${\bf x'}$ in time $t$, composed of straight segments and
undergoing $r$ reflections at the walls.  
For rectilinear geometries, like the parallelepiped that we will discuss, 
this is exact and effectively generalizes the method of images to the Helmholtz equation.

For definiteness consider a scalar field satisfying either Dirichlet or Neumann boundary conditions, introducing a parameter $\eta$, which is $-1$ for the Dirichlet and $+1$ for the Neumann case. The Green's function is then given by
\begin{equation}
G({\bf x}',{\bf x},k)=\sum_{p_r}\frac{\phi(p_r,\eta)}{4\pi l_{p_r}({\bf x}',{\bf x})}e^{ikl_{p_r}({\bf x}',{\bf x})},
\label{flatgreen}
\end{equation}
where $l_{p_r}$ is the length of the path from ${\bf x}$ to ${\bf x'}$ along $p_r$.  
There is a phase factor $\phi(p_r,\eta)=\eta^{n_s+n_c}$ with $n_s$ and $n_c$ the
number of surface and corner reflections, respectively.  
Note that reflections from an \emph{edge} do not contribute to the phase.

Since paths without reflections or with only one reflection can have zero length, 
they require a frequency cutoff $\Lambda$, implemented
by the smoothing function $S_{\Lambda}(k)=e^{-k/\Lambda}$. Then the ${\bf x}$ and $k$
integrals can be exchanged, the $k$ integral performed and the Casimir energy written as
\begin{equation}
E_{\eta}(\Lambda)=\frac{1}{2\pi^2}\sum_{p_{r}}\phi(p_r,\eta)
\int_{\mathcal{D}}d\vx\frac{\Lambda^4(3-(l_{p_r}(\vx)\Lambda)^2)}
{(1+(l_{p_r}(\vx)\Lambda)^2)^3}.
\label{endiv}
\end{equation}
The limit $\Lambda\to\infty$ can be taken in each term of the sum, unless a path has zero length,
which can occur only for cases with $r=0$ or $r=1$.
After isolating these two contributions, we set
\begin{equation}
E_{\eta}(\Lambda)=E_0(\Lambda)+E_1(\Lambda,\eta)+\widetilde{E}_{\eta}.
\end{equation}

The zero reflection path has exactly zero length, and  contributes the energy 
$E_0(\Lambda)=\frac{3}{2\pi^2}V\Lambda^4$, where $V$ is the volume of the space. 
This is a constant and therefore does not contribute to the Casimir force. 
The one reflection paths (energy $E_1(\Lambda,\eta)$) generate cutoff dependent terms, but generically, also cutoff-independent terms. We will show however that such one reflection terms do not contribute to the force when 
specialized to the piston geometry.

For paths undergoing multiple reflections $r>1$, the length $l_{p_r}$ is always finite,
and we can safely send $\Lambda\to\infty$
in eq.~(\ref{endiv}), resulting in the simpler and cutoff independent contribution
\begin{equation}
\widetilde{E}_{\eta}=-\frac{1}{2\pi^2}\sum_{p_{r>1}}\phi(p_r,\eta)\int_{\mathcal{D}}d\vx\frac{1}{l_{p_r}(\vx)^4}.
\label{enfin}
\end{equation}
This is a finite contribution to the energy in the limit $\Lambda\to\infty$. 
The derivative of $\widetilde{E}_{\eta}$ gives the finite force between the rigid bodies.

\subsection{Electromagnetic field modes}
\label{eandm}

In the previous section we defined the optical approach for a scalar field. Although a similar definition can be made for the electromagnetic field in an arbitrary geometry, the Helmholtz equation becomes  matrix--valued, complicating the treatment even in a semiclassical approximation. However, in the piston 
geometry, with arbitrary cross section, the EM field can be separated into transverse magnetic (TM) and transverse electric (TE) modes, that satisfy Dirichlet and Neumann boundary conditions.

At the surface of an ideal conductor, the ${\bf E}$ and ${\bf B}$ fields satisfy the boundary conditions,
 ${\bf E}\times{\bf n}={\bf 0}$ and ${\bf B}\cdot{\bf n}=0$, where ${\bf n}$ is the normal vector at the surface.  The normal modes of the piston consist of a TM set, which satisfy
\begin{equation}E_x=\psi(y,z)\cos\left(n \pi x/a\right),\,\,\,\, n=0,1,2,\cdots,
\label{tm}
\end{equation}
where $\psi$ vanishes on the boundaries of the domain, and therefore satisfies 
Dirichlet conditions on the 2-dimensional boundary;
and a TE set, with
\begin{equation}B_x=\phi(y,z)\sin\left(n \pi x/a\right),\,\,\,\, n=1,2,3,\cdots,
\label{te}
\end{equation}
where $\phi$ satisfies Neumann boundary conditions.    
The other components of ${\bf E}$ and ${\bf B}$ can be computed from Maxwell's equations, and are easily shown to obey conducting boundary conditions.
There is, however, one important exception: 
the TE mode built from the trivial Neumann solution, $\phi={\rm constant}$, does not satisfy 
conducting boundary conditions
unless the constant (and all components of ${\bf E}$ and ${\bf B}$) are zero.
We must ensure that the corresponding set of modes in eq.~(\ref{te}) are not included in
the Casimir summation.

Equations~(\ref{tm}) and (\ref{te}) enable us to list the spectrum of the electromagnetic field.
Denote the spectra of the TM modes as the set $\Omega(N_I\otimes D_\mathcal{S})\subset \mathbb{R}$.
Here $N_I$ indicates that the component on the interval satisfies Neumann boundary conditions,
and $D_\mathcal{S}$ indicates that the component on the cross section satisfies Dirichlet boundary conditions.
Similarly, we denote the spectra of the TE modes as $\Omega(D_I\otimes N_\mathcal{S})$
in a similar notation.  However, as explained above, we must remove $\Omega(D_I)$,
which are the frequencies  with $\phi={\rm constant}$.
Hence, the electromagnetic spectra is the set
\begin{equation}
\Omega_C=\Omega(N_I\otimes D_\mathcal{S}) \cup \Omega(D_I\otimes N_\mathcal{S}) \setminus \Omega(D_I).
\end{equation}
Note that $\Omega(D_I)=\{\pi/a, 2\pi/a, \ldots\}$ is the set of eigenfrequencies in 1-dimension.  The Dirichlet and Neumann spectra on the interval are identical except for the $n=0$ mode 
(see eqs.~(\ref{tm}) and (\ref{te})), but the energy of this mode is independent of $a$ and does not contribute to the Casimir force.  So we may replace $N_{I}
\to D_{I}$ in the TM spectrum and $D_{I}\to N_{I}$ in the TE spectrum, with the result
\begin{equation}
\Omega_C\approx\Omega(D_I\otimes D_\mathcal{S}) \cup \Omega(N_I\otimes N_\mathcal{S}) \setminus \Omega(D_I),
\label{omega}\end{equation}
where the notation $\approx$ indicates equality up to terms independent of $a$.
Thus, the EM spectrum is the union of Dirichlet and Neumann spectra in the 3-dimensional domain, ${\cal D}$, except that the Dirichlet spectrum on the interval must be taken out. 

\section{Rectangular Piston}
\label{rectangle}
\subsection{Derivation}

\begin{figure}[t] 
\begin{center}
\includegraphics[width=\columnwidth]{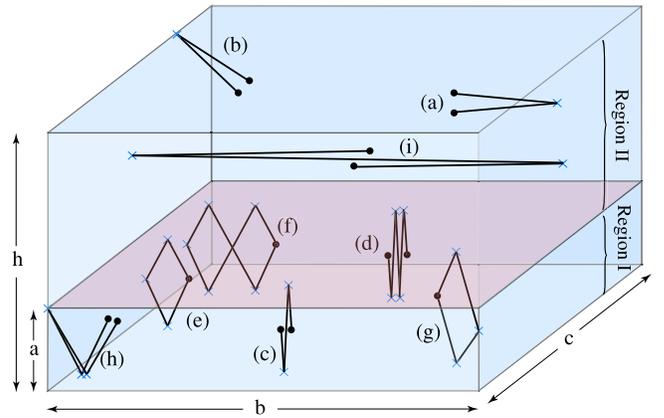}
\caption{\label{picpist}(color online). The 3-dimensional piston of size $h\times b\times c$. A partition at height $a$
separates it into Region I and Region II. 
A selection of representative paths are given in (a)--(i). 
Several of these paths (namely, (a,b,c,d,h,i)) have start and end points that actually coincide,
but we have slightly separated them for clarity.}
\end{center}
\end{figure}

The piston geometry  is depicted in  Fig.~\ref{picpist}. The domain ${\cal D}$ consists of the whole parallelepiped, the union of Regions I and II.  Only the partition, located a distance $a$ from the base and $h-a$ from the top, is free to move.  We study the scalar field for both Dirichlet and Neumann boundary conditions and the electromagnetic field.  According to eq.~(\ref{omega}), the EM Casimir energy arises from the sum of the Dirichlet and Neumann energies in 3 dimensions minus the Dirichlet Casimir energy in one dimension, $E=h\Lambda^2/2\pi-\zeta(2)/(4\pi a)-\zeta(2)/(4\pi (h-a)) $ (a standard result).  In total, then, the EM Casimir force on the partition is 
\begin{equation}
F_C=F_D+F_N+\frac{\zeta(2)}{4\pi a^2}-\frac{\zeta(2)}{4\pi(h-a)^2},
\label{EMf}
\end{equation}
where the final term vanishes if we take $h\to\infty$.

Let us initially focus on Region I, the  parallelepiped of size $a \times b\times c$, below the partition.
The optical energy receives contributions from the sum over all closed paths $p_r$ in domain $\mathcal{D}_{I}$:
Each path is composed of straight segments with equal angles of incidence and reflection 
when bouncing off the walls. 
There are four distinct classes of paths:
$E_{\mbox{\tiny{per}}}$, from periodic orbits reflecting off faces  ({\it {\it e.g.}\/}  paths (c), (d), (i)); 
$E_{\mbox{\tiny{aper}}}$, from aperiodic tours off faces ({\it {\it e.g.}\/}  paths (a), (e), (f));  
$E_{\mbox{\tiny{edge}}}$, from
closed paths involving reflections off edges  ({\it {\it e.g.}\/}  paths (b), (g)); 
and $E_{\mbox{\tiny{cnr}}}$, from closed paths with reflections off corners  ({\it {\it e.g.}\/}  path (h)).
To each path $p_r$, we associate a vector ${\bf l}_{p_r}$ pointed along the {\it initial} heading
of the path, and of length $|{\bf l}_{p_r}|=l_{p_r}$.

First we consider the periodic orbits, which are paths that involve an even number of
reflections off faces, with $r=\{0, 2, 4,\ldots\}$ ({\it {\it e.g.}\/}  paths (c), (d), (i)).
As the central point is varied throughout $\mathcal{D}_{I}$, the length $l_{p_r}$ of each periodic
path remains fixed, making the integration trivial, {\it i.e.\/} $\int_{\mathcal{D}_{I}}d^3x\to abc= V$.
We index the paths by integers $n,m,l$, so ${\bf l}_{p_r}=(2na,\,2mb,\,2lc)$, with $l_{nml}=\sqrt{(2na)^2+(2mb)^2+(2lc)^2}$.
The $n=m=l=0$ term gives $E_0=\frac{3}{2\pi^2}V\Lambda^4$ 
(see eq.~(\ref{endiv}) with $l_{p_{r}}=0$), 
while all others are evaluated using eq.~(\ref{enfin}), giving:
\begin{eqnarray}
E_{\mbox{\tiny{per}}}^I(\Lambda)&=& \frac{3}{2\pi^2}V\Lambda^4 
-\frac{abc}{32\pi^2}Z_3(a,b,c;4)\label{Eper}\\
   &=& \frac{3}{2\pi^2}V\Lambda^4 -\frac{\zeta(4)}{16\pi^2}\frac{A}{a^3}
  +\Gamma(b,c) \nonumber\\
&&  +\mathcal{O}\!\left(e^{-2\pi g/a}\right),\,\,\mbox{as}\,\,a\to 0 \label{Eperapp}
\end{eqnarray}
where $Z_d(a_1,\ldots,a_d;s)$ is the Epstein zeta function defined in the Appendix (eq.~(\ref{Epsteindefn})), and $\Gamma(b,c)$ does not depend on $a$ and hence does not contribute to the force on the piston.
In eq.~(\ref{Eperapp}) $g\equiv\mbox{min}(b,c)$, and $A=bc$ is the area of the base.
The leading cutoff independent piece as $a\to 0$ is the Casimir energy for parallel plates,
coming from orbits that reflect off both the base and partition,
see paths (c), (d), {\it etc.} in Fig.~\ref{picpist}.
To extract this behavior we have used
\begin{equation}
Z_{d}(a_1,\ldots,a_d;s)=\frac{2\zeta(s)}{a_1^s}+\mathcal{O}\left(\frac{1}{a_1}\right),
\label{zeta}\end{equation}
in the regime $a_{1}\ll a_{2},\ldots,a_{d}$ (see Appendix).

We next consider the contribution of the aperiodic orbits 
that involve an odd number of reflections off faces,
with $r=\{1,3,5,\ldots\}$.   Examples in the figure include paths (a), (e), and (f).
For each such path, when we vary the point of integration over $\mathcal{D}_{I}$ one of the Cartesian components of the length vector ${\bf l}_{p_r}$ changes and the other two components are fixed.
For example, only the $x$ component varies for those paths that undergo an odd number of reflections off walls
parallel to the $yz$-plane and an even number
of reflections off walls parallel to both the $xy$ and $xz$-planes.
The $x$-component ${\bf l}_{p_r}$ increases by $2a$ each time that the number of reflections off the $yz$-planes increases, so that ${\bf l}_{p_r}=(2a(n-1)+2\xi(x),\,2bm,\,2cl)$,
where $\xi(x)=x$ or $\xi(x)=a-x$ depending on the direction of the path.
The summation over $n$ and the $x$-integral $\int_{0}^{a}dx$ together combine to form an integral over $x$
from $-\infty$ to $+\infty$. So we introduce $l_{p_r}(x)=\sqrt{(2 x)^2+(2bm)^2+(2cl)^2}$ in terms of which 
the integration over the fixed components $y$ and $z$ is trivial $\int dydz=bc$, and
the $x$-integration is elementary.
The above example singled out the $x$-component. To include all such paths, the analysis must be repeated for the other two components
under the cyclic interchange of $a$, $b$, and $c$.
Employing eq.~(\ref{endiv}) for $\{n,m\}=\{0,0\}$ and eq.\,(\ref{enfin}) for $\{n,m\}\ne\{0,0\}$
we obtain,
\begin{eqnarray}
E_{\mbox{\tiny{aper}}}^I(\Lambda)&=&\frac{\eta}{8\pi}S\Lambda^3
-\frac{\eta}{64\pi}\Big{(}abZ_2(a,b;3)+acZ_2(a,c;3) \nonumber \\
&&+bc\,Z_2(b,c;3)\Big{)}
                          \label{Eaper}\\
                &=&\frac{\eta}{8\pi}S\Lambda^3-\eta\frac{\zeta(3)}{64\pi}\frac{P}{a^2}
        +\Upsilon(b,c) \nonumber\\
        && +\mathcal{O}\!\left(e^{-2\pi g/a}\right),\,\,\mbox{as}\,\,a\to 0 \label{Eaperapp}
\end{eqnarray}
 where $\Upsilon$ does not depend on $a$, $S=2(ab+ac+bc)=aP+2A$ is the total surface area, and $P=2b+2c$ is the perimeter of the base.
The leading cutoff independent piece as $a\to 0$ comes from paths that reflect once off a side wall
and off both the base and partition, see paths (e), (f), {\it etc.} in Fig.~\ref{picpist}.

Next we calculate the contribution of even reflection paths which intersect an edge of Region I.
Examples include (b) and (g) in Fig.~\ref{picpist}.
In this case it is only the component of ${\bf l}_{p_r}$
{\em parallel} to the edge that remains fixed, while the other 2 components vary as the point of origin varies over ${\cal D}_{I}$.
For example, suppose the reflecting edge is parallel to the $z$-axis.  Then the $z$-integration is trivial, $\int_{0}^{c}dz=c$, and the path vector is a function of $x$ and $y$  given by
${\bf l}_{p_r}=(2a(n-1)+2\xi(x),\,2b(m-1)+2\psi(y),\,2cl)$,
where $\xi(x)=x$ or $\xi(x)=a-x$, $\psi(y)=y$ or $\psi(y)=b-y$, depending on the quadrant that ${\bf l}_{p_r}$
lies in: up or down in $x$, right or left in $y$, respectively.  
In this case we can replace both the summations over $n$ and $m$ and the double integral over $x$ and $y$ by
an integral over the whole $xy$-plane.   This integral is most easily performed in polar co-ordinates,
using a path length that may be written as $l_{p_r}(r)=\sqrt{(2an)^2+(2r)^2}$. 
The contribution to the Casimir energy is found to be
\begin{eqnarray}
E_{\mbox{\tiny{edge}}}^I(\Lambda)&=&\frac{1}{32\pi}L\Lambda^2
-\frac{\zeta(2)}{16\pi}\left(\frac{1}{a}+\frac{1}{b}+\frac{1}{c}\right),
\label{Eedge}\end{eqnarray}
where $L=4(a+b+c)=4a+2P$ is the total perimeter length.
 The cutoff independent piece $\sim 1/a$ comes from paths that reflect once off a side edge and off both the base and partition, see path (g), {\it etc.} in Fig.~\ref{picpist}. 
 
Finally, we consider the paths which reflect off a corner ($E_{\mbox{\tiny{cnr}}}$).
In this case, as the integration variable moves throughout its domain, {\em all} components
of the distance vector ${\bf l}_{p_r}$ vary.
Hence, we can incorporate all such paths by extending our integral over
all space in $x$, $y$ and $z$. This leaves no dependence on the geometry of the parallelepiped (i.e., it is independent
of $a$, $b$, and $c$), and only contributes a constant that is of no interest, which we ignore.

Adding together all these contributions, we obtain the Casimir energy of a scalar field
in Region I as
\begin{equation}
E_\eta^I(\Lambda)= \frac{3}{2\pi^2}V\Lambda^4 +\frac{\eta}{8\pi}S\Lambda^3
+\frac{1}{32\pi}L\Lambda^2+\widetilde{E}^I_\eta,
\label{eq1}
\end{equation}
where $\widetilde{E}_{\eta}^I$ gives the cutoff independent piece,
from eqs.~(\ref{Eper}), (\ref{Eaper}), and (\ref{Eedge}).
We note that the cutoff dependent terms agree with the leading terms obtained by  integrating Balian and Bloch's asymptotic expansion of the density of states \cite{Balian:1970fw}. 

We obtain the Casimir energy for the entire piston by adding to eq.~(\ref{eq1}) the analogous expression for Region II obtained by the replacements: $a\to h-a$, $V\to hA$, $S\to hP+4A$, and $L\to 4h+4P$.   
It is easy to see that after including Region II 
the sum of all cutoff dependent terms is independent of partition height $a$.  Therefore the force on the partition is well defined and finite in the limit $\Lambda\to\infty$.  Also, of course, the contribution to the Casimir energy from the region outside the piston is independent of $a$ and can be ignored entirely. The force on the partition is given by the partial derivative with respect to $a$ of the cutoff independent terms as
\begin{equation}
F_\eta=-\frac{\partial}{\partial a}\left(\widetilde{E}_\eta(a,b,c)+\widetilde{E}_\eta(h-a,b,c)\right),
\end{equation}
where we have defined $\widetilde {E}_{\eta}^{I}=\widetilde{E}_{\eta}(a,b,c)$.

We focus on the $h\to\infty$ limit in which
the expression for the contribution from Region II simplifies.  
Consider the periodic, aperiodic, and edge paths whose cutoff independent
contribution to the energy is given in eqs.~(\ref{Eper}), (\ref{Eaper}), and (\ref{Eedge}).
Replacing $a\to h-a$, taking $h\to\infty$,  and using eq.~(\ref{biga}) of the appendix 
in these equations gives
\begin{eqnarray}
&& \widetilde{E}_{\mbox{\tiny{per}}}^{II}\to-\frac{h-a}{32\pi^2 A}Z_2(b/c,c/b;4),\nonumber\\
&& \widetilde{E}_{\mbox{\tiny{aper}}}^{II}\to
-\eta\frac{h-a}{32\pi}\left(\frac{1}{b^2}+\frac{1}{c^2}\right)\zeta(3),\nonumber\\
&&\widetilde{E}_{\mbox{\tiny{edge}}}^{II}\to 0,
\label{regII}\end{eqnarray}
where we have not reported terms independent of $a$, since they do not affect the force. 
Also, we re-express   the Region I energy  $\widetilde{E}_{\eta}(a,b,c)$
in a fashion that is useful for $a\ll b,c$, using eq.~(\ref{zetaprop}) of the appendix.  The net force on the partition due to quantum fluctuations of the {\em scalar field} is then
%
\begin{eqnarray}
F_{\eta}&\!=&\!-\frac{3\zeta(4)}{16\pi^2}\frac{A}{a^4}-\eta\frac{\zeta(3)}{32\pi}\frac{P}{a^3}-\frac{\zeta(2)}{16\pi a^2}
-\frac{J_\eta(b/c)}{32\pi^2 A} \nonumber\\
&\!+&\!\eta\frac{\pi}{2 a^3}\sum_{m,n=1}^{\infty}n^2\left(K_{0}\left(2\pi m n\,b/a\right)b
+(b\leftrightarrow c)\right) \nonumber\\
&\!+&\!\frac{\pi^2}{32}\frac{A}{a^4}\sum_{m,n}\pri\frac{\coth(f_{mn}(b/a,c/a))}{f_{mn}(b/a,c/a)\sinh^2(f_{mn}(b/a,c/a))},\,\,\,\,\,\,\,\,\,\,\,
\label{force3d}\end{eqnarray}
%
where the primed summation is over $\{m,n\}\in\mathbb{Z}^2\setminus\{0,0\}$
and $K_0$ is the zeroth order modified Bessel function of the second kind.
Here we have defined $J_\eta(x)\equiv Z_2(x^{1/2},x^{-1/2};4)+\pi\eta(x+x^{-1})\zeta(3)$ and $f_{mn}(x,y)\equiv
\pi\sqrt{(mx)^2+(ny)^2}$.
The first four terms of eq.~(\ref{force3d}) dominate for $a\ll b,c$,
while the following terms are exponentially small in this regime. 
The first term arises from periodic orbits reflecting off walls (see eq.~(\ref{Eperapp})), 
the second term from aperiodic tours bouncing off walls (see eq.~(\ref{Eaperapp})),
the third term from reflections off edges (see eq.~(\ref{Eedge})), the fourth term from Region II paths (see eq.~(\ref{regII})).
Note that the infinite series, involving exponentially small terms, is convergent for any $a,~b,~c$.

The electromagnetic case is closely related to the scalar Dirichlet and Neumann cases, which
we discussed in detail in Section~\ref{eandm}.  According to eq.~(\ref{EMf}), the EM Casimir energy
in Region I is related to the Dirichlet energy $E_D$ and the Neumann energy $E_N$ by
\begin{equation}
E_C^I(\Lambda)=E_D^I(\Lambda)+E_N^I(\Lambda)-\sum_{d=a,b,c} E_1(d,\Lambda),
\end{equation}
where $E_1(d,\Lambda)=d\Lambda^2/2\pi-\zeta(2)/(4\pi d)$ is the energy of a scalar field in 1-dimensions obeying Dirichlet boundary conditions in a region of length $d$.  The contribution from Region II follows from replacing $a\to h-a$. Combining previous results, the electromagnetic Casimir force is found to be,
\begin{eqnarray}
F_{C}=&\!-&\!\frac{3\zeta(4)}{8\pi^2}\frac{A}{a^4}+\frac{\zeta(2)}{8\pi a^2}
-\frac{J_C(b/c)}{32\pi^2 A}
+\frac{\pi^2}{16}\nonumber\\
&\!\times&\!\frac{A}{a^4}\sum_{m,n}\pri\frac{\coth(f_{mn}(b/a,c/a))}{f_{mn}(b/a,c/a)\sinh^2(f_{mn}(b/a,c/a))},\,\,\,\,\,\,\,\,\,\,
\label{forceEM}\end{eqnarray}
where $J_C(x)\equiv J_{-1}(x)+J_{+1}(x)=2 Z_2(x^{1/2},x^{-1/2};4)$.

\subsection{Discussion}

Here we address the implications of eqs.~(\ref{force3d}) and (\ref{forceEM}) for the
force on the partition in more detail. 
To begin, we discuss the important issue of attraction versus repulsion.  We are interested in comparing
the force on the partition ($F_{\Gamma}$, where $\Gamma=D,N$ or $C$ for Dirichlet, Neumann, or EM boundary conditions respectively) to the force reported in the literature for a single cavity, 
which we denote $F_{\Gamma,{\rm box}}$ \cite{Wolfram}.  
The latter is obtained by the following prescription:  
calculate the energy in a single rectilinear cavity, drop the cutoff dependent (``divergent'') terms, ignore contributions from the region exterior to the cavity, and differentiate with respect to $a$ to obtain a force.
We emphasize that there is no justification for dropping the cutoff dependent terms, so although we refer to this result, for convenience, as $F_{\rm box}$, it does not apply to the physical case of a rectilinear box.

For the piston geometry, we 
note that the {\it sole} contribution from Region II is the $a$--independent term denoted by $J$. 
In fact this is \emph{the only term that distinguishes} $F$ from $F_{\rm box}$, {\it i.e.\/},
\begin{equation}
F_{\Gamma}=F_{\Gamma,{\rm box}}-J_{\Gamma}(b/c)/(32\pi^2 A).
\end{equation}
Naively, the difference by a constant
may not seem important. Indeed it is not too important 
for small values of the ratio $a/(b,c)$. However it is very important for $a\gtrsim (b,c)$.
\begin{figure}[t]
\begin{center}
\includegraphics[width=\columnwidth]{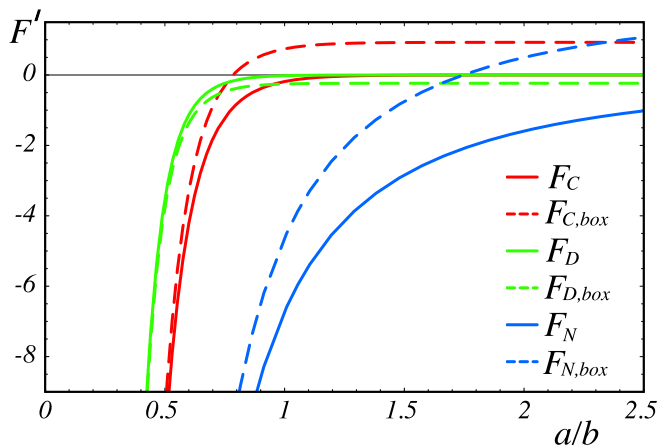}
\caption{\label{unforce}(color online). The force $F$ on a square piston ($b=c$) due to quantum fluctuations of a field
subject to Dirichlet, Neumann, or conducting boundary conditions, 
as a function of $a/b$, rescaled as $F'\equiv 16\pi^2 A F/(3\zeta(4))$ ($F'\equiv 8\pi^2 A F/(3\zeta(4))$)
for scalar (EM) fields. 
The solid lines are for the piston, solid middle = $F_C$, solid upper = $F_D$, and solid lower = $F_N$, while their dashed counterparts are for the box.
 }
\end{center}
\end{figure}
In Fig.~\ref{unforce} we plot both these forces for a square cross section ($b=c$) as a function
of $a/b$. (The plots include scalar as well as EM cases.) Note that in all cases $F\to 0$,
while $F_{\rm box}\to J(1)/(32\pi^2 A)$ (a constant) as $a/b\to\infty$.
For this geometry $J_D(1)\approx -1.5259,\, J_N(1)\approx 13.579$,
and $J_C(1)\approx 12.053$, so $J$ is negative for Dirichlet and positive for both Neumann and EM.
We see that $F$ is always attractive, while $F_{\rm box}$ can change sign.
It is always attractive for Dirichlet, but becomes repulsive
for Neumann when $a/b>1.745$ and for EM when $a/b>0.785$. 
This is the consequence of ignoring Region II and the cutoff dependence.
Indeed, it is easy to show that the piston force is attractive for any choice of $a,b,c,h$.
A final comment is that for $h$ finite and $a=h/2$, the partition sits at an
unstable equilibrium position. This comment was made in Ref.~\cite{Hammer}, although the 
above detailed results were not derived there.

\begin{figure}[t]
\begin{center}
\includegraphics[width=\columnwidth]{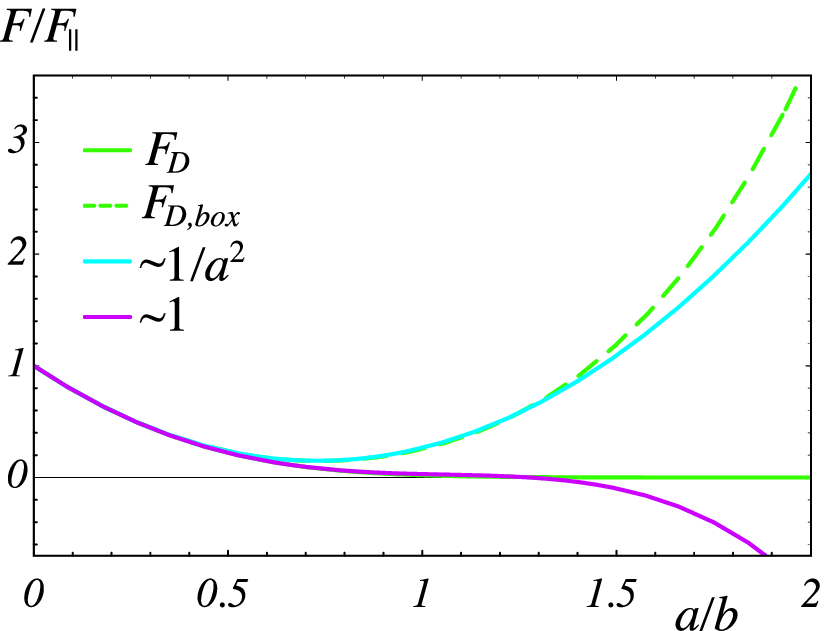}\quad
\includegraphics[width=\columnwidth]{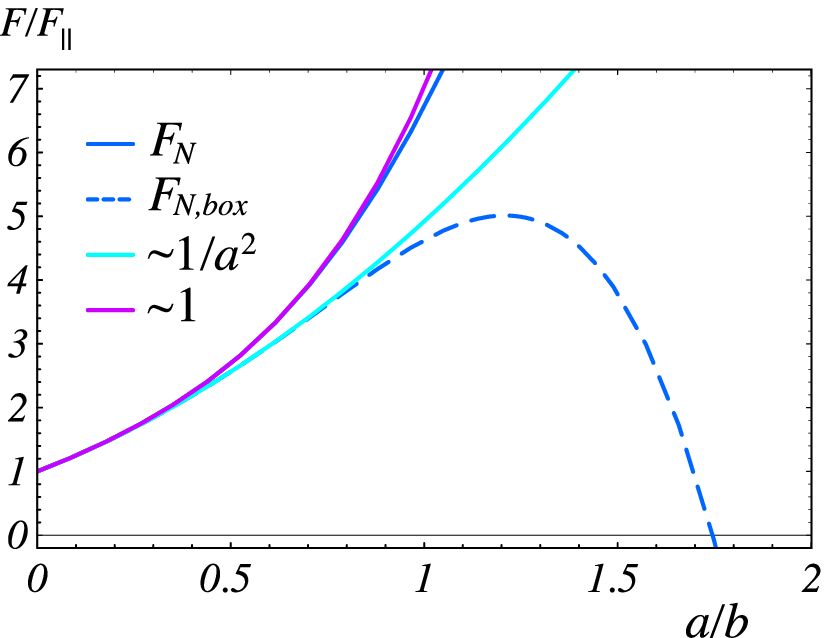}
\caption{\label{picforce}(color online). The force $F$ on a square partition ($b=c$) due to quantum fluctuations of a scalar field 
as a function of $a/b$, 
normalized to the parallel plates force $F_\parallel$. Left figure is Dirichlet; solid middle = $F_D$ (piston), dashed = $F_{D,\mbox{\tiny{box}}}$ (box), solid upper = $\{1/a^4,1/a^3,1/a^2\}$ terms, solid lower = $\{1/a^4,1/a^3,1/a^2,1\}$ terms. Right figure is Neumann; solid middle = $F_N$ (piston), dashed = $F_{N,\mbox{\tiny{box}}}$ (box), solid lower = $\{1/a^4,1/a^3,1/a^2\}$ terms, solid upper = $\{1/a^4,1/a^3,1/a^2,1\}$ terms.}
\end{center}
\end{figure}

\begin{figure}[t]
\begin{center}
\includegraphics[width=\columnwidth]{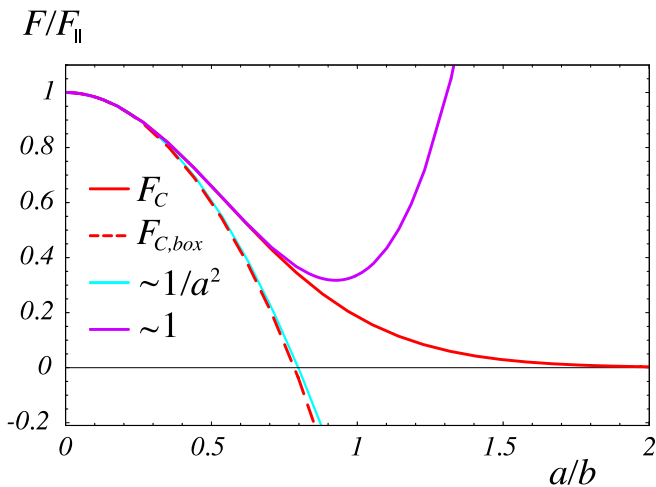}
\caption{\label{picforce2}
(color online). The force $F$ on a square partition ($b=c$) due to quantum fluctuations of the
EM field as a function of $a/b$, 
normalized to the parallel plates force $F_\parallel$. Solid middle = $F_C$ (piston), dashed = $F_{C,\mbox{\tiny{box}}}$ (box), solid lower = $\{1/a^4,1/a^2\}$ terms, solid upper = $\{1/a^4,1/a^2,1\}$ terms.
}
\end{center}
\end{figure}

With the explicit form for $F$, we can more closely compare the piston with Casimir's original parallel plate geometry.  
For better comparison in Figs.~\ref{picforce} and \ref{picforce2}, we have plotted
the forces for the scalar and EM fields, after dividing by the parallel plates results,
$F_{(D,N) \parallel}=-3\zeta(4)A/(16\pi^2 a^4)$ or $F_{C \parallel}=-3\zeta(4)A/(8\pi^2 a^4)$.
First, note that for the EM case
not only does $F_{C}\to 0$ as $a/b\to\infty$ but it does so rather quickly.  
Since $F_{C \parallel}$ vanishes as $1/a^{4}$, it is clear from Fig.~\ref{picforce2} that $F_{C}$ vanishes even more rapidly.
In fact it vanishes exponentially fast, as $e^{-2\pi a/b}$ for $a\gg b$. We can understand this as follows:
In this limit the most important paths are those that reflect off the top and bottom plates once, and therefore travel a distance $2a$. The transverse wavenumber
$k=\pi/b$ due to the finite cross section, acts as an effective mass for the system, and damps the contribution of these paths exponentially. 
In fact for any rectangular cross section we find
\begin{equation}
F_C\approx-\frac{\pi}{2}\left(\frac{1}{\sqrt{a b^3}}e^{-2\pi a/b}
+\frac{1}{\sqrt{a c^3}}e^{-2\pi a/c}\right),\,\,\,\,\mbox{as}\,\,\,a\to\infty.
\end{equation}

Experimentally, values of $a/b\sim 1$ are not yet realizable. 
Instead, typical experimental studies of Casimir forces have transverse dimensions that 
are roughly $100$ times the separation between the ``plates''.
This means that the leading order corrections to $F_{C \parallel}$ are more likely to be detected experimentally.
In Figs.~\ref{picforce} and \ref{picforce2} we show the result of including successive corrections to $F_\parallel$ for scalar and EM cases, respectively; we plot the curve which includes $\{1/a^4, 1/a^3, 1/a^2\}$ terms and another curve that includes $\{1/a^4, 1/a^3, 1/a^2, 1\}$ terms.
In the EM case we note that the $1/a^3$ term that appears
in the expansion for Dirichlet and Neumann boundary conditions is canceled. 
In the next Section, we will demonstrate that this is a general phenomenon for
any cross section (see ahead to eq.~(\ref{FCgen})). 
Hence the  first correction  to the EM result scales as $ 1/a^2$, which is $\mathcal{O}(a^2/A)$ compared
to $F_{C \parallel}$. We see that this correction is quite accurate up to $a/b\sim 0.3$. We suspect this regime of accuracy to be roughly valid for any cross section.  

\section{General cross sections}
\label{general}
The piston for arbitrary cross section cannot be solved exactly, but we can obtain much useful information
from an asymptotic expansion for small separation $a$.
The generalized piston maintains symmetry along the vertical axis, and 
its geometry is the product of $I\otimes {\mathcal S}$ of the interval $I=[0,a]$ and
some 2-dimensional cross section $\mathcal{S}\subset\mathbb{R}^2$.  Let us denote by ${\mathcal E}=k^2$ the eigenvalues of the Laplacian on the piston base ${\mathcal S}$ and the interval $I$, separately, with appropriate boundary conditions
\begin{equation}
-\Delta_{\mathcal{S},I}\psi_{\mathcal{S},I}={\mathcal E}\psi_{\mathcal{S},I}.
\end{equation} 
The corresponding densities of states are denoted by by $\rho_{I}$ and $\rho_{\mathcal{S}}$, respectively. 
Then the density of states (per unit ``energy", ${\cal E}$) of the problem in the 3-dimensional region $I\otimes {\mathcal S}$ is $\rho(\mathcal{E})$ and can be written as the
convolution
\begin{equation}
\rho(\mathcal{E})=\int_{0}^{\infty}d\mathcal{E}'\rho_{\mathcal{S}}(\mathcal{E}
-\mathcal{E}')\rho_{I}(\mathcal{E}').
\end{equation}
The 2-dimensional density $\rho_{\mathcal{S}}$ is not known in general, 
since the wave equation can not be solved in full generality in an arbitrary  domain ${\mathcal S}$.
However, for small height to width ratios, the smallness of $a$ translates to  high energies $\mathcal{E}$,
and we will see that the asymptotic behavior of $\rho_{\mathcal{S}}$ is sufficient for extracting an asymptotic expansion for the force is powers of $1/a$.

The number of eigenstates with energy less than ${\cal E}$ in ${\cal S}$ has the asymptotic 
expansion at large ${\cal E}$ \cite{Baltes},
\begin{equation}
N_{\mathcal{S}}(\mathcal{E})=
\left(\frac{A}{4\pi}\mathcal{E}+\eta\frac{P}{4\pi}\sqrt{\mathcal{E}}+\chi
+r_N(\mathcal{E})\right)\Theta(\mathcal{E}).
\end{equation}
Here, $\chi$ is related to the shape of the domain $\mathcal{S}$ through
\begin{equation}
\chi=\sum_{i}\frac{1}{24}\left(\frac{\pi}{\alpha_i}-\frac{\alpha_i}{\pi}\right)
+\sum_{j}\frac{1}{12\pi}\int_{\gamma_j}\kappa(\gamma_j)d\gamma_j,
\end{equation}
where $\alpha_i$ is the interior angle of each sharp corner and
$\kappa(\gamma_j)$ is the curvature of each smooth section.
It is easy to check that $\chi=1/4$ for a rectangle and $\chi=1/6$ for any smooth shape (for example a circle).
Note that we have included the step function $\Theta({\cal  E})$ in the expression for $N_{\mathcal{S}}({\cal E})$, ensuring
that only $\mathcal{E}>0$ contributes. Here $r_N(\mathcal{E})$ is a function which designates
lower order terms (remainder) in an $\mathcal{E}\to\infty$ asymptotic expansion.
For any polygonal shape $r_N$ is exponentially small, 
$r_N(\mathcal{E})=\mathcal{O}(e^{-c\mathcal{E}})$ ($c>0$ is a constant) \cite{Bailey}. 
However, we are aware of only a much weaker estimate for smooth shapes, as 
$r_N(\mathcal{E})=\mathcal{O}(1/\sqrt{\mathcal{E}})$ \cite{Baltes}.
The derivative of $N_\mathcal{S}({\cal E})$ is the density of states\footnote{\label{dosfootnote}
In Eqs.~(\ref{dos2d}) and (\ref{densexp}) we have denoted the various remainders by 
$r_{\rho}(\mathcal{E}), r_1(\mathcal{E}), r_2(\mathcal{E})$.
We discuss the size of the remainders in the expansion of the forces $F_{\eta}$ \& $F_C$ following eq.~(\ref{FCgen}).}
\begin{equation}
\rho_\mathcal{S}(\mathcal{E})=
\left(\frac{A}{4\pi}+\eta\frac{P}{8\pi}\frac{1}{\sqrt{\mathcal{E}}}\right)\Theta(\mathcal{E})
+\chi\,\delta(\mathcal{E})+r_{\rho}(\mathcal{E}), \label{dos2d}\end{equation}
where we have used $\Theta'(\mathcal{E})=\delta(\mathcal{E})$,
and $\mathcal{E}\,\delta(\mathcal{E})=\sqrt{\mathcal{E}}\,\delta(\mathcal{E})=0$ for all $\mathcal{E}$.

The other function in the convolution, the 1-dimensional density of states, is known exactly: it is simply a sum over delta functions,
which we rewrite in terms of its Poisson summation
\begin{eqnarray}
\rho_I(\mathcal{E})=\sum_{n=1}^{\infty}&\!\!\!\delta&\!\!\!\!\left(\mathcal{E}-\frac{n^2\pi^2}{a^2}\right)\\
=\frac{a}{2\pi}\frac{\Theta(\mathcal{E})}{\sqrt{\mathcal{E}}}&\!\!\!+&\!\!\!
2\sum_{m=1}^{\infty}\int_{0}^{\infty}dx \cos(2\pi m x)\,\delta\!\left(\mathcal{E}-\frac{x^2\pi^2}{a^2}\right)\!.
\,\,\,\,\,\,\,\,\,\,
\label{rho1d}\end{eqnarray}
The first term in eq.~(\ref{rho1d}) is the smooth contribution to the density of states,
and the second is the oscillatory component.
The leading contributions (as $a\to 0$) to the 3-dimensional density of states come from convolving the smooth
part of $\rho_\mathcal{S}$ with $\rho_I$, giving$^{\ref{dosfootnote}}$
\begin{eqnarray}
\rho(\mathcal{E})&=&a\left(\frac{1}{4\pi^2}A\sqrt{\mathcal{E}}+\frac{\eta}{16\pi}P+\frac{1}{2\pi}
\frac{\chi}{\sqrt{\mathcal{E}}}+r_1(\mathcal{E})\right)\nonumber\\
&+&\!\sum_{m=1}^{\infty}\Big{(}\frac{A}{4\pi^2 m}\sin(2ma\sqrt{\mathcal{E}})+\eta\frac{aP}{8\pi}
J_{0}(2ma\sqrt{\mathcal{E}})\nonumber\\
&+&\!\frac{a \chi}{\pi\sqrt{\mathcal{E}}}\cos(2ma\sqrt{\mathcal{E}})+r_2(\mathcal{E})\Big{)}. 
\label{densexp}\end{eqnarray}
The first line agrees precisely with the Balian and Bloch theory of the density of states \cite{Balian:1970fw},
and gives the cutoff dependent terms in the Casimir energy
\begin{equation}
E(\Lambda)=\frac{1}{2}\int_0^\infty d\mathcal{E}\rho(\mathcal{E})\sqrt{\mathcal{E}}\,e^{-\sqrt{\mathcal{E}}/\Lambda}. 
\end{equation}
The cutoff dependent contributions have no effect on the Casimir force when Region II is included, 
since they are linear in $a$, as explained earlier.
The second line in eq.~(\ref{densexp}) 
gives the leading three terms in an asymptotic expansion of the force
\begin{equation}
F_\eta=-\frac{3\zeta(4)}{16\pi^2 a^4}A-\eta\frac{\zeta(3)}{32\pi a^3}P-\frac{\zeta(2)\chi}{4\pi a^2}+r_\eta(a).
\label{Fetagen}\end{equation}
Also, even for these general cross sections, the EM energy can be related using eq.~(\ref{EMf}) 
to Dirichlet and Neumann energies, as
\begin{equation}
F_C=-\frac{3\zeta(4)}{8\pi^2 a^4}A+\frac{\zeta(2)(1-2\chi)}{4\pi a^2}+r_C(a).
\label{FCgen}\end{equation}
In eqs.~(\ref{Fetagen}) and (\ref{FCgen}) we have written the remainder terms
as $r_\eta(a)$ and $r_C(a)$ $(=r_{-1}(a)+r_{+1}(a))$, respectively.
Following our earlier estimates for $r_N(\mathcal{E})$ that appears in $N_{\mathcal{S}}(\mathcal{E})$,
and noting that there is always an $\mathcal{O}(1)$ term that comes from Region II,
we have $r_{\eta,C}(a)=\mathcal{O}(1)$ for polygonal shapes and 
$r_{\eta,C}(a)=\mathcal{O}(1/a)$ for smooth shapes, as $a\to 0$.

The generalization to arbitrary cross sections in eq.~(\ref{FCgen})
has interesting features. 
The correction to the parallel plates result depends on geometry through the parameter $\chi$,
which depends sensitively on whether the cross section is smooth or has sharp corners. 
For example, $\chi=\frac{1}{6}$ for all smooth shapes
and $\chi=\frac{1}{6}\frac{n-1}{n-2}$ for an $n$-sided polygon of equal interior angles.
Given unavoidable imperfections in any experimental realization, one may wonder
what precisely constitutes ``smooth" or ``sharp."
Note that for any deformation with local radius of curvature $R$ ($R=0$ for perfectly sharp corners),
we have the dimensionless quantity $R/a$, where  $a$ is the base--partition height.
Given that our expansion is valid for small $a$, we conclude that $R\gg a$ is a
smooth deformation, while $R\ll a$ can be regarded as a sharp corner.
As a simple example, consider a shape that is roughly square (4-sided polygon) 
if viewed from large distances,
but is in fact rounded with radius $R$ at the ``corners" if examined closely. 
Let us also imagine that the overall width ($b$) is much larger than $R$.
Then, since the corresponding term in the Casimir force goes as $1-2\chi$ (see eq.~(\ref{FCgen})),
we expect the correction to be $\zeta(2)/(6\pi a^2)$ for $a/R\ll 1$ and decrease to 
$\zeta(2)/(8\pi a^2)$ for $a/R\gg 1$.
A more interesting example would be a self-similar (fractal or self-affine) perimeter, in which the number
of sharp corners deceases as a power of the resolution $a$.
For such a case, we expect a correction scaling as a non-trivial power of $1/a$, reminiscent
of results in Ref.~\cite{LiKardar}.
It would be interesting to see if such corrections are experimentally accessible.

Another noteworthy feature of  eq.~(\ref{FCgen}) is that the leading correction to the EM force
(compared to parallel plates) is smaller by order of $a^2/A$. 
By contrast the corrections are only of order $a/\sqrt{A}$ for scalar fields with either Dirichlet or
Neumann boundary conditions.
However, the latter corrections are exactly equal and opposite in sign, and cancel for the EM force. It is interesting to inquire if this precise cancellation applies only to perfect metallic
boundary conditions, or remains when 
the effects of finite conductivity are taken into account.
More work is necessary to understand the finite conducting piston.  Yet another case is for side walls made of dielectrics, where a simple modification of the optical path method, which replaces the sign factor $\eta$ with 
the reflection coefficients for TM and TE modes, suggests that the cancellation 
does not occur. A piston that is made entirely of a uniform dielectric is
examined in Ref.~\cite{Bartondiel}

\section{Thermal corrections}
\label{thermal}
The question of the leading corrections to the Casimir force at finite temperatures $T$ has
generated recent interest, both from the practical need to evaluate the accuracy of experiments,
and due to fundamental issues.
In particular, there is controversy pertaining to the appropriate model for the metallic walls,
which we shall ignore in this chapter.
Instead, we shall compute corrections to the Casimir force due to finite temperature
excitations of the modes in the piston, while continuing to treat its walls as perfect metals \cite{Geyer03}. 

\subsection{Rectangular piston}

We first answer this question for the piston with rectangular cross section.
In units with $\hbar=c=k_B=1$, the inverse temperature $\beta=1/T$  introduces a 
new length scale whose size relative to the dimensions $a$, $b$, and $c$ 
of the piston (we imagine, as earlier, that $h\to\infty$) sets the importance of thermal corrections. 
(More precisely, $\pi\beta$ is the appropriate length scale.)
In typical experiments $a\sim 1\mu m$, $b,c\sim 100\mu m$, 
and at room temperature $\pi\beta\sim 20\mu m$. 
Thus the regime of most experimental interest is where the length scales
satisfy $a\ll\pi\beta\lesssim b,~c$. 
In light of this we focus on thermal lengths much larger than
the base--partition height, {\it i.e.} $a\ll\pi\beta$. 
To fully investigate the low temperature regime, we assume  $a\ll \pi\beta,b,c\ll h$,
but will allow $\pi\beta$ to be less than or greater than $b$ or $c$.

Each mode of the field can be regarded as an independent harmonic oscillator, and by 
summing the corresponding contributions, we find the free energy
\begin{equation}
\mathcal{F_{\rm tot}}=\frac{-1}{\beta}\sum_{m}\ln\left(\frac{e^{-\frac{1}{2}\beta \omega_m}}{1-e^{-\beta\omega_m}}\right)=E+\delta\mathcal{F}.
\end{equation}
We have separated out the the zero-temperature Casimir energy $E$, from the finite temperature
``correction" $\delta\mathcal{F}=\delta E - T\delta S$,
and focus on the latter for calculating finite temperature effects.

First, a note of caution is in order regarding the scalar field with Neumann boundary conditions.
In {\it any cavity}, there is a trivial solution to the Neumann problem, 
namely a constant field with $\omega=0$. 
This means that whenever $\beta$ is finite ($T>0$) then $\delta\mathcal{F}=-\infty$,
which signals condensation of the scalar field into the ground state.
We note that this phenomenon occurs for {\it closed} geometries where the spectrum is discrete
and not in general for {\it open} geometries in which the region near $\omega=0$
is integrable due to phase space suppression. 
We will proceed by calculating the free energy of a scalar field with both Dirichlet  and Neumann 
boundary conditions, ignoring the mode with $\omega=0$ for the latter. 
We then use eq.~(\ref{EMf}) to obtain the EM force.
This procedure is valid since the offending Neumann mode is specifically excluded
from the EM spectrum.
 
 For a Dirichlet scalar field in Region I, since all modes satisfy $\omega_m>\pi/a$,
their Boltzmann weights are small in the limit of $a\ll\pi\beta$, and
\begin{equation}
\delta\mathcal{F}^{I}=\mathcal{O}\!\left(e^{-\pi\beta/a}\right)
\label{expsmall}\end{equation}
is exponentially small. 
Similarly, the $a$--dependent terms of the {\em electromagnetic} free energy in region I
are exponentially small. 
This is true for any cross section and reflects
the fact that thermal wavelengths $\sim\pi\beta$ are {\it excluded} from Region I \cite{Scard2}.
However, a power law contribution to the free energy and force will come from Region II.  
We use the optical expansion, which remains exact for the free energy in rectilinear geometries,  to compute this contribution for scalar fields \cite{Scard2}, as
\begin{equation}
\delta\mathcal{F}^{II}=-\frac{1}{2\pi^2}\sum_{p_r}\phi(p_r,\eta)\sum_{q}\pri \int_{\mathcal{D}}d{\bf x}
\frac{1}{\left[l_{p_r}({\bf x})^2+(q\beta)^2\right]^2}.
\end{equation}
Note that here the sum ranges over $q\in\mathbb{Z}\setminus\{0\}$ ---
the $q=0$ term is just the Casimir energy (see eq.~(\ref{enfin})).

It is natural to break the energy up into the familiar four classes of paths:
periodic orbits, aperiodic tours off faces, reflections off edges,
and reflections off corners. However, summing each set separately gives a logarithmic divergence 
(that cancels among the different classes for Dirichlet boundary conditions ).
Fortunately, this problem can be ignored in the $h\to\infty$ limit, as can be seen,
for example, by considering the contribution  from the sum over periodic orbits (paths (c), (d), etc in Fig.~1).
Noting that $h-a$ is the height of the piston in Region II, we have 
\begin{eqnarray}
\delta\mathcal{F}^{II}_{\mbox{\tiny{per}}}=&\!\!-&\!\!\frac{1}{16\pi^2}\sum_{q=1}^{\infty}\sum_{n,m,l=-\infty}^{\infty} \nonumber\\
&\!\!\times&\!\!\frac{(h-a)bc}{\left[(n(h-a))^2+(mb)^2+(lc)^2+(q\beta/2)^2\right]^2}.\,\,\,\,\,\,\,\,\,\,
\end{eqnarray}
This expression is logarithmically divergent, but if we take $h\to\infty$, only the $n = 0$ term contributes and the remaining summation over $\{q, m, l\}$ is finite. 
Strictly speaking, the interchange of the limit $h\to\infty$ with the summations, which eliminates the logarithmic divergence, is formally problematic. 
However a more rigorous analysis justifies this step for the Dirichlet case through the cancellation
among the different classes, but always gives $-\infty$ for the Neumann case as anticipated.
Performing this interchange gives the following result for the contribution of periodic orbits
\begin{eqnarray}
\delta\mathcal{F}^{II}_{\mbox{\tiny{per}}}=
&\!\!-&\!\!\frac{\zeta(4)(V_p - Aa)}{\pi^2\beta^4}-\frac{(h-a)A}{32\pi\beta}Z_2(b,c;3)
\nonumber\\
&\!\!+&\!\!\frac{(h-a)A}{32\pi^2}Z_2(b,c;4)+\mathcal{O}\!\left(e^{-4 \pi g/\beta}\right)
\end{eqnarray}
with $g\equiv\min(b,c)$ and $V_p$ as the total piston volume.   Here we have expanded for small $\beta$ relative to $g=\min(b,c)$.
We note that although the third term is independent of $\beta$, this really is part
of $\delta\mathcal{F}$. The reader that is interested in the opposite limit of $\beta\to\infty$, i.e., the low temperature limit, should look ahead to Section \ref{LowTemp}. 

Proceeding in a similar fashion for all contributions to the free energy of a scalar field
we find (ignoring the exponentially small contribution of Region I)
\begin{eqnarray}
\delta\mathcal{F}_{\eta}=&-&\frac{\zeta(4)(V_p-Aa)}{\pi^2\beta^4}
-\eta\frac{\zeta(3)(S_p-Pa)}{8\pi\beta^3}\nonumber\\
&-&\frac{\zeta(2)(h-a)}{4\pi\beta^2}
-\frac{M_{\eta}(b/c)(h-a)}{32\pi\beta\sqrt{A}}\nonumber\\
&+& \frac{J_{\eta}(b/c)(h-a)}{32\pi^2 A} -\frac{\pi^2(V_p-Aa)}{8\beta^4}\nonumber \\
&\times &
\sum_{m,n}\pri\frac{\,1+2f_{mn}(\bar{b},\bar{c}) - e^{-f_{mn}(\bar{b},\bar{c})}}
{f_{mn}^{\,3}(\bar{b},\bar{c})\mbox{sinh}^2(f_{mn}(\bar{b},\bar{c}))}
-\eta\frac{(h-a)}{\beta^2}\nonumber\\
&\times & \sum_{m,n=1}^{\infty}\frac{n}{m}\left(K_1(2\pi m n \bar{b})+K_1(2\pi m n\bar{c})\right)
\label{freeD}
\end{eqnarray}
where we have defined $M_{\eta}(x)\equiv Z_2(x^{3/2},x^{-3/2};3)+4\eta(x^{1/2}+x^{-1/2})\zeta(2)$,
$\bar{b}\equiv 2b/\beta, \bar{c}\equiv 2c/\beta$, and $S_p$ is the total surface area of the piston.
It is important to note that while $\delta\mathcal{F}_{-1}=\delta\mathcal{F}_{D}$, 
$\delta\mathcal{F}_{+1}=\delta\mathcal{F}_{N}$ is not strictly correct as we have ignored the $\omega=0$ Neumann mode. 
Although $\delta\mathcal{F}_{N}=-\infty$, as stated earlier, this expression correctly gives the $a$-dependence 
in $\delta\mathcal{F}_N$.

The EM case can be handled in a similar fashion.
Repeating our earlier decomposition, we note that 
$\delta\mathcal{F}_{\mbox{\tiny{EM}}}=\delta\mathcal{F}_{-1}+\delta\mathcal{F}_{+1}+\zeta(2)(h-a)/\beta^2$, since the spectral decomposition in eq.~(\ref{omega}) {\it correctly} leaves out the $\omega=0$ mode
of the Neumann spectrum.
We thus find (again ignoring the exponentially small contribution of Region I)
\begin{eqnarray}
\delta\mathcal{F}_{\mbox{\tiny{EM}}}&\!\!=&\!\!-\frac{2\zeta(4)(V_p-Aa)}{\pi^2\beta^4}
+\frac{\zeta(2)(h-a)}{2\pi\beta^2} \nonumber\\
&\!\!-&\!\!\frac{M_C(b/c)(h-a)}{32\pi\beta\sqrt{A}}
+\frac{J_C(b/c)(h-a)}{32\pi^2 A} \nonumber\\
&\!\!-&\!\!\frac{\pi^2(V_p-Aa)}{4\beta^4}
\sum_{m,n}\pri\frac{\,1+2 f_{mn}(\bar{b},\bar{c}) - e^{-2 f_{mn}(\bar{b},\bar{c})}}
{f_{mn}^{\,3}(\bar{b},\bar{c})\mbox{sinh}^2(f_{mn}(\bar{b},\bar{c}))}\,\,\,\,\,\,\,\,\,\,
\label{freeEM}
\end{eqnarray}
where $M_C(x)\equiv M_{-1}(x)+M_{+1}(x)=2 Z_2(x^{3/2},x^{-3/2};3)$.

In Eqs.~(\ref{freeD}) and (\ref{freeEM}) we have written the expansion as a series 
in increasing powers of $\beta$. The result, though, is correct (up to exponentially small terms in $a/\pi\beta$) for any ratio of $\beta$ to $b$ or $c$, and for  
$h$ much larger than any of the other scales. 
The infinite summations that appear are convergent for all finite values of $\{\beta,b,c\}$.
The leading term in eq.~(\ref{freeEM}) is the Stefan--Boltzmann energy, and the following terms are
corrections due to geometry. 
The term independent of $\beta$ is equal to but opposite in sign to that appearing in the Casimir energy.
 Note that the appearance of a term independent of $\beta$ is an artifact of performing 
a small $\beta$ expansion. 
All terms depend linearly on $a$ and provide a constant force on the partition.
Note that the first five terms in $\delta\mathcal{F}_{\eta}$
and the first four terms in $\delta\mathcal{F}_{\mbox{\tiny{EM}}}$ have power law dependences on $\beta$,
while the remaining terms (summations) are exponentially small for $\pi\beta < (b,c)$.

\subsection{General cross section}

If we consider general $I \otimes \mathcal{S}$ geometries, as in Section~\ref{general}, we may use the smooth 3-dimensional Balian and Bloch density of states in Region II to obtain the leading terms in the free energy.  
Specifically, we use the first line of eq.~(\ref{densexp}) with the replacement $a\to h-a$ for $\rho(\mathcal{E})$,
and calculate the free energy from
\begin{equation}
\delta\mathcal{F}=\frac{1}{\beta}\int_{0}^{\infty}d\mathcal{E}\,
\rho(\mathcal{E})\ln(1-\exp(-\beta\sqrt{\mathcal{E}})). 
\end{equation}
Since we only know the first three terms in the expansion  for the density of states,
we will obtain contributions proportional to the volume, surface, and perimeter of the piston,
but nothing at $\mathcal{O}(1/\beta)$. It is fairly straightforward to get
\begin{eqnarray}
\delta\mathcal{F}_{D}&=&-\frac{\zeta(4)(V_p-Aa)}{\pi^2\beta^4}
+\frac{\zeta(3)(S_p-Pa)}{8\pi\beta^3} \nonumber\\
&&-\frac{\zeta(2)\chi(h-a)}{\pi\beta^2}+\mathcal{O}\!\left(\frac{1}{\beta}\right),\\
\delta\mathcal{F}_{\mbox{\tiny{EM}}}&=&-\frac{2\zeta(4)(V_p-Aa)}{\pi^2\beta^4}
+\frac{\zeta(2)(1-2\chi)(h-a)}{\pi\beta^2} \nonumber\\
&&+\mathcal{O}\!\left(\frac{1}{\beta}\right).
\label{FTgen}\end{eqnarray}
We again see the effect of the modes excluded from Region I due to $a\ll \pi\beta$,
in the factors $V_p-Aa$, $S_p-Pa$, and $h-a$.
These leading terms provide  thermal contributions to the quantum force on the partition, 
given in eqs.~(\ref{Fetagen}) and (\ref{FCgen}).

Let us comment on the relationship between the Casimir and thermal contributions to the force.
We begin by focusing on the regime that is perhaps of most experimental interest: $a\ll\pi\beta\ll\sqrt{A}$.
If we include both Casimir and thermal contributions to the force, as given in eqs.~(\ref{FCgen}) \& (\ref{FTgen}), 
\begin{eqnarray}
F_{\mbox{\tiny{EM}}}=&-&\frac{3\zeta(4)}{8\pi^2}\left(\frac{1}{a^4}+\frac{1}{(\beta/2)^4}\right)A
\nonumber\\
&+&\frac{\zeta(2)(1-2\chi)}{4\pi}\left(\frac{1}{a^2}+\frac{1}{(\beta/2)^2}\right)+\cdots. \,\,\,\,
\label{both}\end{eqnarray}
Note that the leading contributions are related to terms in the Casimir energy by the interchange
$a\leftrightarrow\beta/2$, but this connection ceases for higher order corrections.
We have only calculated further terms for the parallelepiped and we can compare them in this limit.
In particular,  eq.~(\ref{freeEM})  includes a contribution of order $1/\beta$ which has no
 counterpart ({\it i.e.\/} a term of order $1/a$) in the Casimir force.
A term of order $1/a$ can only come from the derivative of  $\sim\ln a$, 
which is absent from the EM Casimir energy.

\subsection{Low temperature limit}\label{LowTemp}

Equation~(\ref{both}) provides the leading terms in the Casimir force in the limit $a\ll \pi\beta\ll b,c$
(or more generally $a\ll \pi\beta\ll\sqrt{A}$ for non-rectangular cross sections).
We may more accurately refer to this as a ``medium temperature'' regime,
as opposite to a lower temperature regime with $\pi\beta\gg\sqrt{A}$. 
In fact, for the rectangular piston we obtained in  eqs.~(\ref{freeD}) and (\ref{freeEM})
results that are valid for $a\ll\{\pi\beta,b,c\}$ for any ratio of $\beta$ to $b$ or $c$,
and will now examine their lower temperature limit.
A naive application of the proximity-force approximation gives always a thermal correction to the force that vanishes as $\sim 1/\beta^4=T^4$ in the $T\to 0$ limit \cite{group}. 
However, in Ref.~\cite{Scard2} it is argued that for open geometries this limit is quite subtle and is sensitive to the detailed shape of each surface. In fact it is reasonable to argue that for the cases relevant to experiments there may be weaker power laws, {\it i.e.\/}, $1/\beta^\alpha$ with $\alpha<4$. 
But in our closed geometry another scenario is natural: If $T\to 0$, so that $\beta\gg a,\sqrt{A}$, modes are excluded from {\it both} regions due to a gap in the spectrum, resulting
in an exponentially small free energy, which (for the rectangular piston) is
\begin{eqnarray}
\delta\mathcal{F}_{\mbox{\tiny{EM}}}&=&-\frac{(h-a)}{\sqrt{2}\beta^{3/2}}
\left(\frac{1}{\sqrt{b}}e^{-\pi\beta/b}+\frac{1}{\sqrt{c}}e^{-\pi\beta/c}\right),\nonumber\\
&& \mbox{as}\,\,\,\beta\to\infty.
\label{highbeta}\end{eqnarray}

\begin{figure}
\includegraphics[width=\columnwidth]{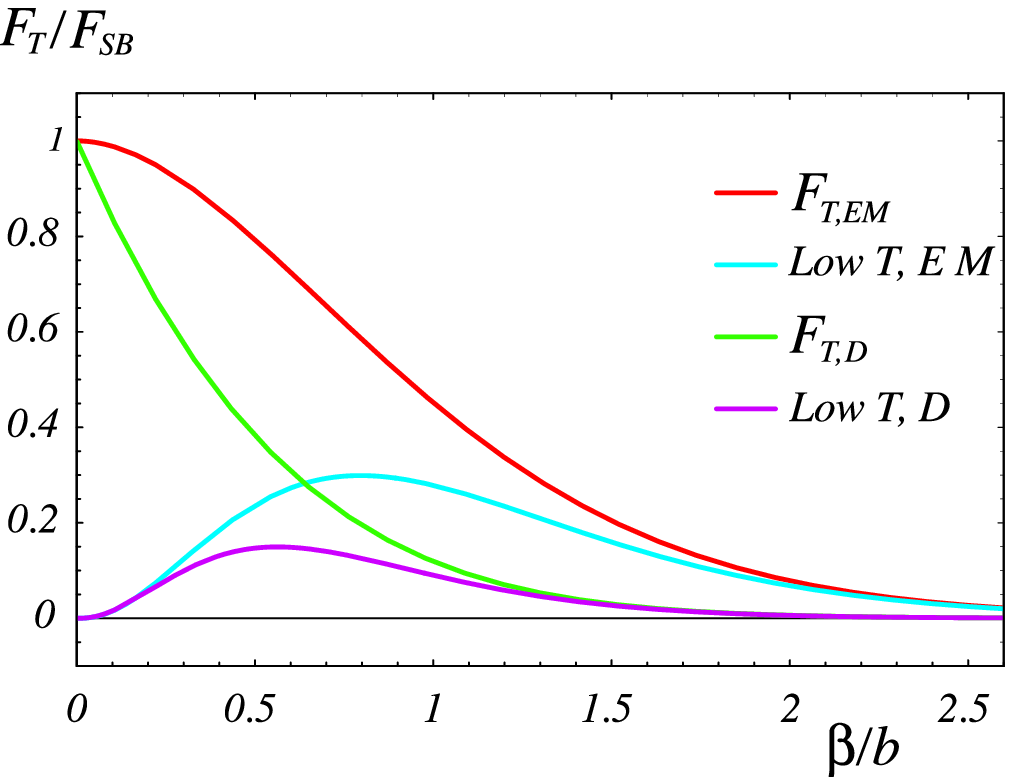}
\caption{\label{thermalplot}(color online). The force $F_T$ from thermal fluctuations on a square partition ($b=c$), normalized
to the Stefan--Boltzmann expression $F_{SB}=-\zeta(4)A/(\pi^2\beta^4)$ ($-2\zeta(4)A/(\pi^2\beta^4)$)
for Dirichlet (EM) fields, as a function of $\beta/b$. This is valid in the regime: 
$a\ll \{\pi\beta, b,c\}$. 
Starting from a normalized value of 1, the full result for Dirichlet (electromagnetic) is the lower (upper) curve. Also, starting from a normalized value of 0, the exponentially small asymptote (as $\beta/b\to\infty$) for Dirichlet (electromagnetic) is the lower (upper) curve.
} 
\end{figure}

A plot of the force $F_{T\,\Gamma} \equiv-d\delta{\cal F}_{\Gamma}/da$ (where $\Gamma=D$ or $C$ as for $T=0$), derived from eqs.~(\ref{freeD}) and
(\ref{freeEM}) is given in Fig.~\ref{thermalplot}. The
force is normalized to the Stefan--Boltzmann term,
$F_{SB}=-\zeta(4)A/(\pi^2\beta^4)$ ($-2\zeta(4)A/(\pi^2\beta^4)$)
for Dirichlet (EM) fields. Having taken $a\ll\pi\beta$ in our analysis, the $a$ dependence is ignorable,
and we plot the force as a function of $\beta/b$ ($b=c$). 
The high $\beta/b$ asymptotic curves (eq.~(\ref{highbeta}) is the EM case) are also included. 
Note that from eq.~(\ref{both}) we can read off the small $\beta/\sqrt{A}$
corrections to $F_{SB}$ for arbitrary cross sections.

\section{Conclusions}
\label{conclusions}

In this work we have obtained an exact, analytic result for the Casimir force for a piston geometry.
Exact, analytic results are rare in this field but nonetheless particularly useful for comparison with the approximations needed to describe real systems and more complicated geometries.

We have obtained analytic expressions for the force acting on the partition in a piston with perfect 
metallic boundaries.
The results are exact for the rectangular piston, and in the form of an asymptotic series in $1/a$ for arbitrary cross section.
We find that the partition is always attracted to the (closer) base;
consistent with a more general result obtained in Ref.~\cite{Klich}.
Since the piston geometry is closely related to single cavity for which a repulsive force has been conjectured,
we are able to shed some light on this question.
In particular, we emphasize that to avoid unphysical deformations (and closely related issues
on cutoffs and divergences) it is essential to examine contributions to the force from both
sides of the partition. The cutoff independent contribution from a single cavity (that we call $F_{\mbox{\tiny{box}}}$) approaches a constant for large $a$.
However, in the piston geometry compensating contributions from the second cavity 
cancel both the cutoff dependent terms and part of the cutoff independent term,
to cause a net attraction.

For general cross sections we find interesting dependence on geometrical
features of the shape, such as its sharp corners and curved segments.
We have obtained the first three terms for scalar fields and the first two terms
for EM fields (one less due to cancellation) 
in an expansion in powers of $a$. By
comparison to our calculated exact result for a rectangular cross section we estimate that
this expansion is valid for $a/b\approx 0.3$. This covers the conventional
experimentally accessible regime, and is therefore a useful result for a large class of geometries. 
We have also obtained thermal corrections which cover the experimentally accessible regime.

\vspace*{2ex}
\section*{Acknowledgments}
We thank M.~Schaden.
M.~P.~H., R.~L.~J., and A.~S. are supported in part by funds provided by the U.S.~Department of
Energy (D.O.E.) under cooperative research agreement DE-FC02-94ER40818.
M.~K. is supported by NSF grant  DMR-04-26677. 

\vspace*{2ex}
\section*{APPENDIX}
The general Epstein Zeta function is defined as
\begin{equation}
Z_d(a_1,\ldots,a_d;s)\equiv \sum_{n_1,\ldots,n_d}\prii\left((n_1 a_1)^2+\cdots + 
(n_d a_d)^2\right)^{-s/2},
\label{Epsteindefn}\end{equation}
where the summation is over $\{n_1,\ldots,n_d\}\in\mathbb{Z}^d\setminus \{0,\ldots,0\}$.
Note that the Riemann Zeta function is a special case of this, namely $\zeta(s)=Z_1(1;s)/2$.

In eq.~(\ref{zeta}) we pointed out that such functions could be approximated
by a term involving the Riemann zeta function and a power of $a_1$, as $a_1\to 0$.
An exact representation, as discussed in Ref.~\cite{Wolfram}, is 
\begin{eqnarray}
&& \!\!\!\!\!\!\!\!\!\!\! Z_{d}(a_1,\ldots,a_d;s) \nonumber \\
& = & \frac{2\zeta(s)}{a_{1}^s}
+\frac{\Gamma\!\left(\frac{s-1}{2}\right)\sqrt{\pi}}{\Gamma\!\left(\frac{s}{2}\right)a_1}
Z_{d-1}(a_2,\ldots,a_d;s-1)\nonumber\\
&+&\frac{4\pi^{s/2}}{\Gamma\!\left(\frac{s}{2}\right)a_{1}^{s}}\sum_{n=1}^{\infty}
\sum_{n_2,\ldots,n_d}\prii n^{(s-1)/2} \nonumber\\
&&\times\, K_{(s-1)/2}\left(2\pi n\frac{\sqrt{(a_2n_2)^2+\cdots
+(a_d n_d)^2}}{a_1}\right)\nonumber\\ 
&& \times\left(\frac{\sqrt{(a_2n_2)^2+\cdots +(a_dn_d)^2}}{a_1}\right)^{(1-s)/2}
\label{zetaprop}\end{eqnarray}
where $K_{\nu}$ is the modified Bessel function of the second kind. 
This is useful in Region I
where $a_1$ is small (with $a_1\to a$).

For Region II it is important to examine the limit in which
one of the lengths is infinite, say $a_1\to\infty$ (with $a_1\to h-a$). In this limit the order of the zeta function is reduced:
\begin{equation}
Z_{d}(a_1,\ldots,a_d;s) \to Z_{d-1}(a_2,\ldots,a_d;s).
\label{biga}\end{equation}

\end{document}